\documentclass[aps,prb,twocolumn,float,a4paper,amsmath,amssymb]{revtex4}
\newcommand{\be}{\begin{equation}}
\newcommand{\ee}{\end{equation}}
\newcommand{\bea}{\begin{eqnarray}}
\newcommand{\eea}{\end{eqnarray}}

\usepackage[dvips]{color}
\definecolor{g-blue}{rgb}{0.83,0.95,1}
\definecolor{g-yellow}{rgb}{1,1,0.7}
\definecolor{g-green}{rgb}{0.9,1,0.9}
\definecolor{green}{rgb}{0,0.6,0}
\definecolor{cyan}{rgb}{0,0.7,0.7}
\definecolor{black}{rgb}{0,0,0}
\definecolor{grey}{rgb}{0.4 ,0.4 ,0.4 }

 \usepackage{graphicx}
 \usepackage{amsmath,bm,epsfig}
\usepackage{verbatim}
\def \ed {\end{document}}
\def\Fbox#1{\vskip1ex\hbox to 8.5cm{\hfil\fboxsep0.3cm\fbox{%
  \parbox{8.0cm}{#1}}\hfil}\vskip1ex\noindent}  

\def\be{\begin{equation}}\def\ee{\end{equation}}
\def\bea{\begin{eqnarray}}\def\eea{\end{eqnarray}}
\def\bse{\begin{subequations}}\def\ese{\end{subequations}}
\newcommand{\BE}[1]{\begin{equation}\label{#1}}
\newcommand{\BEA}[1]{\begin{eqnarray}\label{#1}}
\newcommand{\BSE}[1]{\begin{subequations}\label{#1}}

\let \= \equiv \let\*\cdot \let\~\widetilde \let\^\widehat \let\-\overline

  \def\1{\bm1} 

\def\<{\left\langle}    \def\>{\right\rangle}
\def\({\left(}          \def\){\right)}
 \def \[ {\left [} \def \] {\right ]}

\newcommand{\C}[1]{{\mathcal{#1}}}    

\renewcommand{\sb}[1]{_{\text {#1}}}  
\def\Sb#1{_{\scriptscriptstyle\rm{#1}}}

\begin{document}

\title{  Comment on
``Symmetry of Kelvin-wave dynamics and the Kelvin-wave cascade in the
$T = 0$ superfluid turbulence''
}
\author{Victor S. L'vov$^*$ and  Sergey V. Nazarenko$^\dag$}
\affiliation{$^*$Department of Chemical Physics,  Weizmann Institute  of Science, Rehovot 76100, Israel \\
$^\dag$University of Warwick, Mathematics Institute
Coventry, CV4 7AL, UK}

\begin{abstract}
We comment on the paper by E. B. Sonin,   PRB \textbf{85}, 104516 (2012)  with which we find ourselves in serious disagreement. We use this option to shed light on some important issues of a theory of Kelvin wave turbulence, touched in E. B. Sonin's paper, in particular,  on the relation between the Vinen spectrum of strong-  and the L'vov-Nazarenko spectrum  of weak-turbulence of Kelvin waves. We also discuss the role of  our  explicit calculation of the Kelvin wave interaction Hamiltonian and of the ``symmetry arguments" that  are supposed to resolve  a contradiction between the Kozik-Svistunov and L'vov-Nazarenko spectrum of weak turbulence of Kelvin waves.
\end{abstract}

\date{\today}

\maketitle
 \subsection*{Introduction}
    Because of its importance in superfluid turbulence and of the
growing experimental capabilities in this field,  there has
recently been a renewed interest in the statistical physics of
Kelvin waves propagating on a vortex line.  A complete understanding
of the statistical behaviour of Kelvin waves is therefore
crucial in order to develop a theory of superfluid turbulence. There were various attempts to
find the energy spectrum of Kelvin wave turbulence. In historical ordering they are
\begin{subequations}\label{Ek}
\begin{eqnarray}\label{V}
E\Sb{V}(k)&\propto& \epsilon^{0}~~~k^{-1}\,, ~\  \mbox{Vinen}~\cite{Vinen};\\ \label{KS}
E\Sb{KS}(k)&\propto& \epsilon^{1/5}k^{-7/5}\,,\ \mbox{Kosik--Svistunov, (KS)}~\cite{KS};\\ \label{LN}
E\Sb{LN}(k)&\propto&\epsilon^{1/3} k^{-5/3}\,,\  \mbox{L'vov--Nazarenko (LN)}~\cite{LN}.~~~
\end{eqnarray}
\end{subequations}
Here $\epsilon$ is the $k$-independent energy flux in the ``inertial interval" of wave vectors $k$ located between the energy pumping scale, $k\sb{in}$, and the dissipation scale, $k\sb{dis}$: $k\sb{in}\ll k\ll k\sb{dis}$.

The Vinen spectrum~\eqref{V} describes \emph{strong wave turbulence}, when the inclination angle $\varphi$ of vortex lines  away  from straight lines is not small, $\varphi\sim 1$.    In this case,  as it is well known in the theory of wave turbulence~\cite{ZLF,Naz,L}, a step-by-step cascading of the energy is absent, the energy flux over scales $\epsilon(k)$ is not constant and thus becomes irrelevant.   The power spectrum of strong wave turbulence
 (as a rule) is determined by the structure of singularities in physical space. Well known examples are surface waves on deep water:  when the acceleration at the top of the water waves exceeds the gravity acceleration $g$ there
are  discontinuities  of the first derivative of the wave profile (creation of ``white
horses''). In $k$-representation this corresponds to the universal
\begin{subequations}\label{SW-F}
\begin{eqnarray}
E(k)&\simeq& \frac{\rho g}{k^3}\,, \  \mbox{Phillips  spectrum of   gravity waves;}~~~\\
\label{SW-H}
E(k)&\simeq& \frac{\sigma}{k}\,, \ \mbox{Hix  spectrum of   capillary    waves.}
\label{SW-H}
\end{eqnarray}
\end{subequations}
Here $\rho$ is the fluid density and $\sigma$ is the surface tension. As it is expected, all the spectra~\eqref{V} and \eqref{SW-F}   are independent of the irrelevant  parameter $\epsilon$. They can be found from dimensional reasoning, using the facts, that for gravity waves
the only remaining parameter in the problem is the gravity acceleration $g$, for the capillary waves -- surface tension $\sigma$,  and for the Kelvin waves, as Vinen realized, this is   the circulation quantum $\kappa$. Physically speaking,  the spectrum~\eqref{V} is a consequence of the vortex reconnections,   that (presumably) happen for  at all  scales and  lead to creation of  {discontinuities} of the vortex directions.

 Philips, Hix and Vinen spectra of\emph{ strong wave turbulence}, being  independent of $\epsilon$,
belong to the same class of so-called {\em critical balance} states in which the linear and the nonlinear time scales are balanced for each $k$.
It was explained in book \cite{Naz} that
the critical balance states arise due to a wave strength limiting process, eg. wave breaking of water waves or reconnections of Kelvin waves.

The KS and LN spectra~\eqref{KS} and \eqref{LN} are related to the \emph{weak wave turbulence} of Kelvin waves, in which the angle $\varphi$ is assumed to be small. In the theory of weak wave turbulence, the Hamiltonian of wave interaction can be expanded in series of small wave amplitudes (inclination angle for the  {Kelvin waves and for the gravity and capillary surface waves}) and only the first nontrivial term in this expansion, describing the interaction of $p\ge 3$ waves determines the turbulent energy spectra.  For surface capillary waves, $p=3$, for the surface gravity waves (in which $3$-wave processes are forbidden by the conservation laws~\cite{ZLF}) $p=4$, for the Kelvin waves $p=6$, as correctly was found by KS~\cite{KS}.

Within the framework of   wave turbulence~\cite{ZLF,Naz} the energy flux over scales, $\epsilon$,  is proportional to the wave-collision integral, St$_p(k)$ which, in its turn is proportional to the energies of $p-1$ waves $E(k')$, $E(k^{''})\,,\dots E(k^{(p-1)})$ participating in a $p$-wave ``collision" processes.  Under assumption of locality of the energy transfer, when the leading contribution to St$_p(k)$ comes from $k', k^{''}, \dots k^{(p-1)} \sim k$ one immediately concludes that $\epsilon \propto [E(k)]^{p-1}$ or
\begin{equation}\label{p-dep}
E(k)\propto \epsilon^{1/(p-1)}\ .
\end{equation}
These simple arguments one can be found in books~\cite{ZLF,Naz} or, e.g. in Online Lecture Course~\cite{L}.

KS spectrum~\eqref{KS} was found in~\cite{KS} as a result of $3  \Leftrightarrow3 $
 Kelvin wave scattering ($p=6$) under assumption of the interaction locality. However in~\cite{LLNR}, the
locality assumption used in \cite{KS} was checked and
shown to be violated for  $3  \Leftrightarrow3 $ Kelvin-wave interactions.
This invalidated the local theory, and a nonlocal theory was
proposed~\cite{LN} resulting in $1  \Leftrightarrow3 $ Kelvin-wave interactions (with $p=4$). This
has prompted a lively debate about the correct spectrum of
Kelvin waves in \cite{KS1,LL,KS2,LLN} which was summarized in two Abu Dhabi workshops on superfluid turbulence in May 2011~\cite{AbuDhabi} and June 2012.

The bottom line of these discussions  is very simple: the basic assumptions and the calculation schemes were the same in both KS~\cite{KS} and LN~\cite{LN} approaches. Namely, the initial Hamiltonian formulation of the Biot-Savart equation of the vortex line motion was the same,  the Hamiltonian expansion approach up to the six-order terms (under assumption of smallness of the Kelvin wave amplitudes) was identical, the canonical transformation technique aiming at the elimination of the four-order terms was the same and only the results were different. In Ref.~\cite{LLNR} we presented an explicit infrared (IR) asymptotic form of the effective 6-wave interaction amplitude,
\begin{equation}\label{W}
\C W_{1,2,3}^{4,5,6}= - \frac{3}{4\pi} k_1k_2k_3k_4k_5k_6\,,
\end{equation}
which directly leads to LN spectrum~\eqref{LN}.

To encourage our colleagues to check our derivation, we have made it publicly  available in the form of a line-by-line commented {\em Mathematica} code ~\cite{calc}. On the other hand,
 KS have not calculated the IR asymptotic  of $\C W_{1,2,3}^{4,5,6}$. Instead,  they presented an argument based on symmetry considerations whose aim  was to show
at showing that $\C W_{1,2,3}^{4,5,6}$ cannot have linear IR asymptotics and, therefore, cannot have the form (\ref{W}). In the other words, KS claimed that \emph{our derivation of Eq.~\eqref{W} contains algebraic mistakes}.
The symmetry argument of KS was refuted in \cite{LL,LLN} and in further discussions at the 2011 and 2012 Abu Dhabi workshops; It was shown that the presence of tilt symmetry does not imply the absence of linear IR asymptotics
in the nonlinear interaction coefficients. Thus, the resolution of the controversy must be done by a careful rigorous derivation rather than by further hand-waving avoiding the direct check. This was summed up in the
2011 Abu Dhabi workshop by LN by a call to ``put your Hamiltonian on the table!"

Unfortunately, we cannot say that the paper by Sonin~\cite{Sonin}  clarifies  the issue. Besides more or less simple statements with which we agree, it has a    set of unclear, questionable and sometimes even incorrect hand-waving arguments.  These arguments  are  related to two main issues in the theory of Kelvin wave turbulence in $T = 0$ superfluids:
 \begin{enumerate}
 \item The role of the tilt symmetry in Kelvin wave turbulence;
   \item The properties of strong Kelvin wave turbulence and of the Vinen spectrum~\eqref{V}.
  \end{enumerate}
  Because of their importance  we found it  timely to further spell out  our position to the superfluid physics community  and to comment on at least some of Sonin's statements.  In particular, we will clarify several issues and past results which appear to have been misinterpreted in Ref.~\cite{Sonin}.

Sonin~\cite{Sonin} has also suggested an alternative scenario of the crossover between the classical and the quantum regions of energy spectra  {in a way that doesn't involve a} bottleneck energy accumulation near the inter-vortex scale $\ell$.  This problem is closely related  {with that of the} zero-temperature limit of  the effective Vinen viscosity $\nu'(0)$, which is much smaller then its value $\nu'(T)$ for $T\lesssim T_\lambda$~\cite{Golov}.    {Explanation of this effect was suggested in our bottleneck papers~\cite{LNR} and   not addressed in the Sonin paper~\cite{Sonin}.} We disagree with the Sonin scenario and are going to return to this question soon during discussion more general problem: temperature dependence of $\nu'(T)$ in wide temperature range from $T\to 0$ to $T\to T_\lambda$.

\subsection*{Role of the tilt symmetry in Kelvin wave turbulence}
It is known that the tilt symmetry of the Kelvin wave hamiltonian is broken when the nonlinearity is truncated at a finite order of the wave amplitude. Sonin in Sec.~II of Ref.~\cite{Sonin} illustrates his points by
considering a tilt transformation of an exact fully nonlinear solution.   Within the
Local Induction Approximation the frequency of this exact Kelvin wave solution is
\begin{equation}\label{fr}
\omega= \frac{ \kappa \Lambda k^2}{4 \pi \sqrt{1+a^2k^2}}\ .
\end{equation}
Here $a$ is the wave amplitude, $\Lambda =\ln (\ell/a_0)$ where $a_0$ is the vortex core radius.
\paragraph{Sonin's first objection:}
\begin{description}
 \item
 \emph{``The mechanism of L'vov et al.~\cite{LN}  is absent in the coordinate frame, with the axis coinciding with the average position of the vortex line in which average vortex displacement and tilt are absent."}
\item And later:
\item \emph{``The mechanism of L'vov et al. originates from the quasistatic Kelvin mode, which in the limit of small $k$ is equivalent to a tilt $\varphi \sim ka$ of the $z$ axis. The tilt can be removed by transformation   to another coordinate frame, in which the mechanism disappears."}

\end{description}
This is incorrect. The scenario of L'vov et al. is not eliminated
by introducing the frame with the axis coinciding with the average position of the vortex line, because it involves the {\em rms} tilt at and near the forcing scales and not the straight average of the line tilt. The forcing scale is the largest scale for the direct cascade setup, but it is of course much less than the total system size. Thus, the forcing scale motions make many oscillations within the containing ``box" and they cannot be eliminated by  {any} rotation  {of any} angle. For simplicity, one can consider an idealized system where the scales in between the box size and the forcing scale are damped (suppressed inverse cascade setup). This is precisely the frame with zero mean tilt that, if it exists, is relevant for the Wave Turbulence construction. When the average tilt does not stabilize at a finite value in the infinite box limit then the average should simply be taken over finite distances which are considerably greater than the forcing scale.

\paragraph{Second Sonin's objection:}
\begin{description}
 \item \emph{``Since the main contribution to Kelvin-wave dynamics comes from the sixth-order terms, the Hamiltonian used by L'vov et al. violates tilt symmetry".}
\end{description}
 So what?   What Lebedev, L'vov and Nazarenko actually showed (in responses~\cite{LL,LLN}  to Kozik and Svistunov objections) is that the tilt symmetry does not prevent the interaction coefficients {\em of any order} to   have linear
in $k$ asymptotics. This is completely different from the claim that the truncated system is tilt symmetric which is, of course, wrong but which has never been made  {in the first place}.
Moreover, the fact that all the higher order terms are needed does not contradict the fact that only the leading order (six-wave) nonlinear terms are important for the wave turbulence spectrum when the forcing is weak.

\paragraph{One more of Sonin's objections:}
\begin{description}
 \item \emph{``In summary, tilting of the axis affects distribution in $k$ space.}\\
\emph{Notice that Lebedev et al.\cite{LL} revealed evidence of this effect in the series expansion in $k$ space and called it the nonlinear shift of the Kelvin-wave frequency with the wave of small $k$. They argued that this was a nontrivial observable physical effect, which supported their position. Without arguing the observability of the effect, I would prefer to call it a visual rather than a physical effect, which has nothing to do with the global symmetry at the border."}
\end{description}
 This is also incorrect. The nonlinear frequency shift has nothing to do with  {any sort of} redistribution in the $k$-space but rather it is a change  {of} frequency, at a fixed $k$, with respect to the frequency of the linear waves. It is clearly present in E. B. Sonin's example of  {a} monochromatic Kelvin wave with  {a} frequency given by formula~\eqref{fr} above. Note that,  {to} leading order, the frequency correction is $-\frac 1 {2} k^4 a^2$ which confirms the linear  asymptotics of the four-mode coupling  coefficient with respect to each of the four coupled wavevectors;  {They} are all  {asymptotically} equal to $k$ in this example. This is a real observable effect!\\

 {We think that at the time  being  it would be reasonable to postpone discussion of other similar objections by Sonin  to later time. Right now, we would only like to candidly} repeat our May-2011  {appeal} ``\emph{put your Hamiltonian on the table!}". Up to now we {haven't yet received a} response from KS. That is why we may only hope that  Sonin will finalize the discussion either by presenting his own public  {available} line by-line calculation of the explicit form of the interaction amplitude $\C W$ or by  {kindly pointing out exactly} in which line {/lines} of our calculation~\cite{calc} we made, according to his opinion, a mistake.

\subsection*{I. Strong Kelvin wave turbulence and Vinen spectrum}

In Sec.~IV of Ref.~\cite{Sonin} Sonin considers  the case when the nonlinearity parameter $\varphi$ is not small and we are dealing with the \emph{strong wave turbulence} (without saying this explicitly).  {In this case}, as we explained above, one cannot  expand   the interaction Hamiltonian.  {The} step-by-step cascading of the energy is absent, the energy flux over scales $\epsilon(k)\ne$const.  and thus  becomes irrelevant  together with the hypothesis of the interaction locality.

 {Sonin's discussion of this problem provides an illustration of how invalid arguments can sometimes combine in ways that eventually produce the correct answer:   Vinen-2003 spectrum~\eqref{V} of strong Kelvin wave turbulence~\cite{Vinen}.}  First he made a strange statement  that in the formal   expansion of the interaction Hamiltonian (without small parameter, $\varphi\sim 1$) ``\emph{the higher-order terms can be important as well, or even more important"}. Next he assumed the interaction locality for any $p$, which is highly questionable and (as we believe) even wrong for $p=6$. Then he applied Eq.~\eqref{p-dep} from the theory of weak wave turbulence, which is not valid in the case of strong turbulence. After that he  {took the limit} $p\to \infty$ in Eq.~\eqref{p-dep} and came to correct conclusion that the energy spectra of strong wave turbulence is independent of the (irrelevant) parameter~$\epsilon(k)$.

A positive content of Sec.~IV in Sonins' paper~\cite{Sonin} is  {that he was able to erase, in implicit way, the} important question of the relations of the strong- and weak-wave turbulence regimes of Kelvin-wave turbulence, i.e., as we believe, between the Vinen and L'vov-Nazarenko spectra~\eqref{V} and \eqref{LN}. We will return to this issue elsewhere.

\emph{\textbf{Acknowledgements}}. We are grateful Anna Pomyalov, Laurent Bou\`e and Eduard Sonin for useful discussions of this Comments. This work is supported by the EU FP7 Microkelvin program  (Project No.~228464) and by  the U.S. Israel Binational Science Foundation.

\end{document}